\documentclass{iaus}
\usepackage{graphicx}

\title[Multiple stellar populations in star clusters] 
{Observations of multiple populations in star clusters}

\author[G. Piotto]   
{Giampaolo Piotto$^1$
}

\affiliation{$^1$Dipartimento di Astronomia, Universit\`a di Padova,
  \\ Vicolo dell'Osservatorio, 3, I-35122, Padova, Italy \\ email: {\tt giampaolo.piotto@unipd.it} }

\pubyear{2009}
\volume{258}  
\pagerange{1--10}
\jname{The Ages of Stars}
\editors{}
\begin{document}

\maketitle

\begin{abstract}
An increasing number of photometric observations of multiple stellar
populations in Galactic globular clusters is seriously challenging the
paradigm of GCs hosting single, simple stellar populations. These
multiple populations manifest themselves in a split of different
evolutionary sequences as observed in the cluster color-magnitude
diagrams. Multiple stellar populations have been identified in Galactic and Magellanic Cloud clusters. In this paper we will summarize the observational scenario.

\end{abstract}

\firstsection 

\section{Introduction}

Globular star clusters (GC) have occupied a prominent role in our 
understanding of the structure and evolution of (low mass) stars. At 
the basis of the use of GCs as templates for stellar models was the
assumption that their stars can be idealized as "simple stellar
populations" (SSP), i.e. as an assembly of coeval, initially chemically
homogeneous, single stars.
Thanks to this idea, GCs, and star clusters in general, have been used
for decades to test and calibrate synthetic models
of stellar populations, a critical tool for studying galaxies at low
as well as at high redshift.  

Color-magnitude diagrams (CMD) like the magnificent CMD of NGC 6397 by
Richer et al. (2008, see also Anderson et al. 2008)
fully support the paradigma of GCs hosting simple stellar
populations.  However, there is a growing body of observational facts
which challenge this traditional view.  Since the eighties we know
that GCs show a peculiar pattern in their chemical abundances (see
Gratton el al. 2004 for a recent review). While they are generally
homogeneous insofar Fe-peak elements are considered, they often exhibit
large anticorrelations between the abundances of C and N, Na and O, Mg
and Al. These anticorrelations are attributed to the presence at the
stellar surfaces of a fraction of the GC stars of material which have
undergone H burning at temperatures of a few ten millions K (Prantzos
et al. 2007; less for the C and N anticorrelation). This pattern is
peculiar to GC stars; field stars only show changes in C and N
abundances expected from typical evolution of low mass stars (Sweigart \& Mengel 1979); it is
primordial, since it is observed in stars at all evolutionary phases
(Gratton et al. 2001); and the whole stars are interested (Cohen et
al. 2002).


In addition, since the sixties (Sandage and Wildey 1967, van den Bergh 1967), we know that the horizontal branches (HB) of some GCs can be
rather peculiar.  This problem, usually
known as the {\it the second parameter} problem, still lacks of a
comprehensive understanding: many mechanisms, and many parameters have
been proposed to explain the HB peculiarities, but none apparently is
able to explain the entire observational scenario. It is well
possible that a combination of parameters is responsible for the HB
morphology (Fusi Pecci et al. 1993). Surely, the total cluster mass
seems to have a relevant role (Recio-Blanco et al. 2006).
\begin{figure}[h!]
\begin{center}
 \includegraphics[width=4.0in]{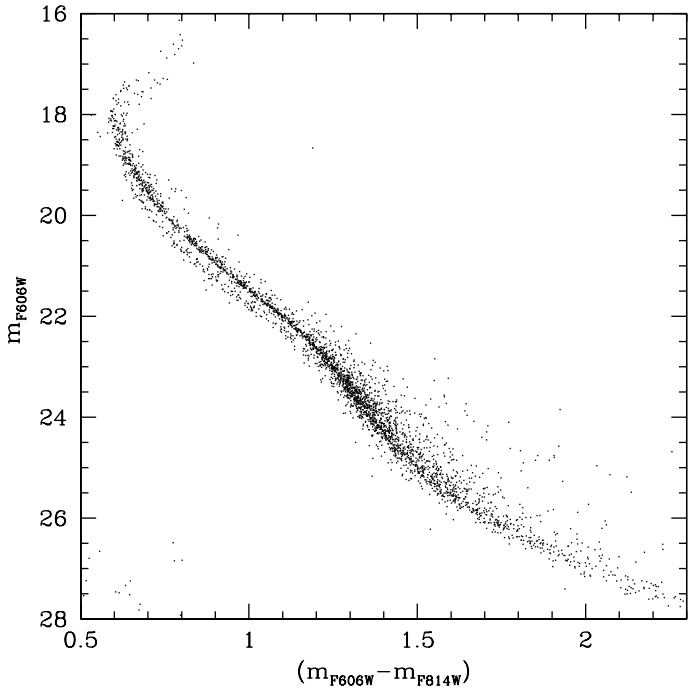} 
 \caption{The multiple MS of Omega Centauri. This spectacular 
  CMD (from Bedin et al. 2009, in preparation) 
  comes from multiepoch observations of a field at 17 arcmin
  from the cluster center. The plotted stars have been selected on the
  basis of their proper motion and all of them are cluster members.
  The split of the two MSs is clearly visible from the TO down to 
  $m_{F606W}=23.0$. Then the two MS seems to merge, due to the increased
  photometric error. Some stars are on the red side of the two main MSs; they
  are too far from the red MS to be binaries. These stars correspond to the 
  third MS discussed in Villanova et al. (2007), which is likely related to the 
  RGB-a of Pancino et al. (2002).}
   \label{fig1}
\end{center}
\end{figure}

It is tempting to relate the second parameter problem to the complex
abundance pattern of GCs. Since high Na and low O abundances are
signatures of material processed through hot H-burning, they should be
accompanied by high He-contents (D'Antona \& Caloi 2004). In most
cases, small He excesses up to $\delta Y\sim0.04$ (that is Y~0.28, assuming the
original He content was the Big Bang one) are expected. While this
should have small impact on colors and magnitudes of stars up to the
tip of the RGB, a large impact is expected on the colors of the HB
stars, since He-rich stars should be less massive. E.g., in the case
of GCs of intermediate metallicity ([Fe/H]~-1.5), the progeny of
He-rich, Na-rich, O-poor RGB stars should reside on the blue part of
the HB, while that of the "normal" He-poor, Na-poor, O-rich stars
should be within the instability strip or redder than it. Actually
mean HB colors are influenced by the mass loss along the RGB and by
small age differences of 2-3
Gyr. However, within a single GC a correlation is expected between the
distribution of masses (i.e. colors) of the HB-stars and of Na and O
abundances.

In summary, a number of apparently independent observational facts
seems to suggest that, at least in some GCs, there are stars which
have formed from material which must have been processed by a previous
generation of stars.

The questions is: do we have some direct, observational evidence of
the presence of multiple populations in GCs? Very recent discoveries,
made possible by high accuracy photometry on deep HST images, allowed
us to positively answer to this question. In this paper, we will
summarize these new observational facts, and briefly discuss their
link to the complex abundance pattern and to the anomalous HBs.

\section{Observational Evidence of Multiple Populations in GCs}

\subsection{The first discoveries}

\begin{figure}
\centering
\resizebox{6.5cm}{!}{\includegraphics{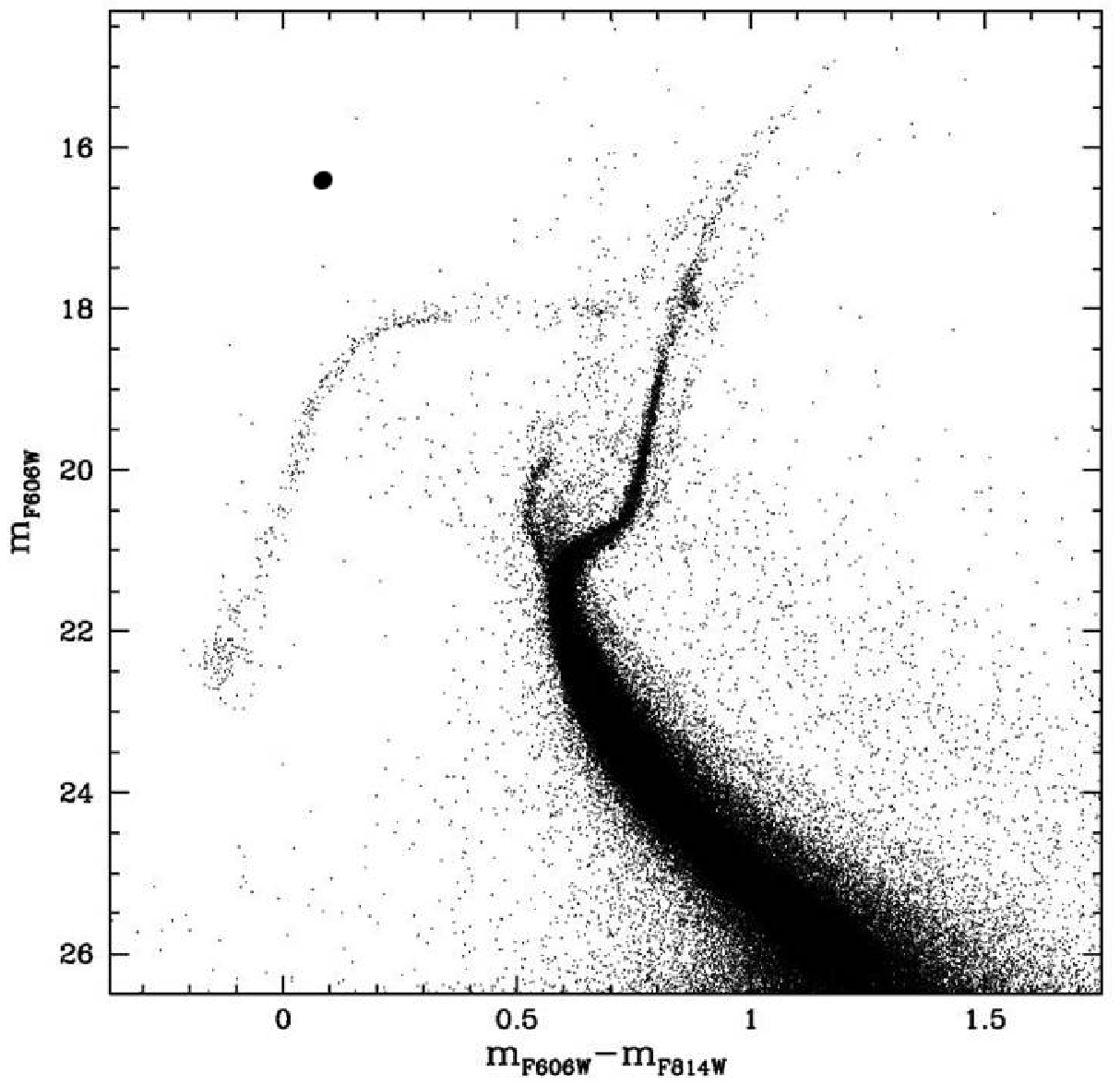} }
\resizebox{6.5cm}{!}{\includegraphics{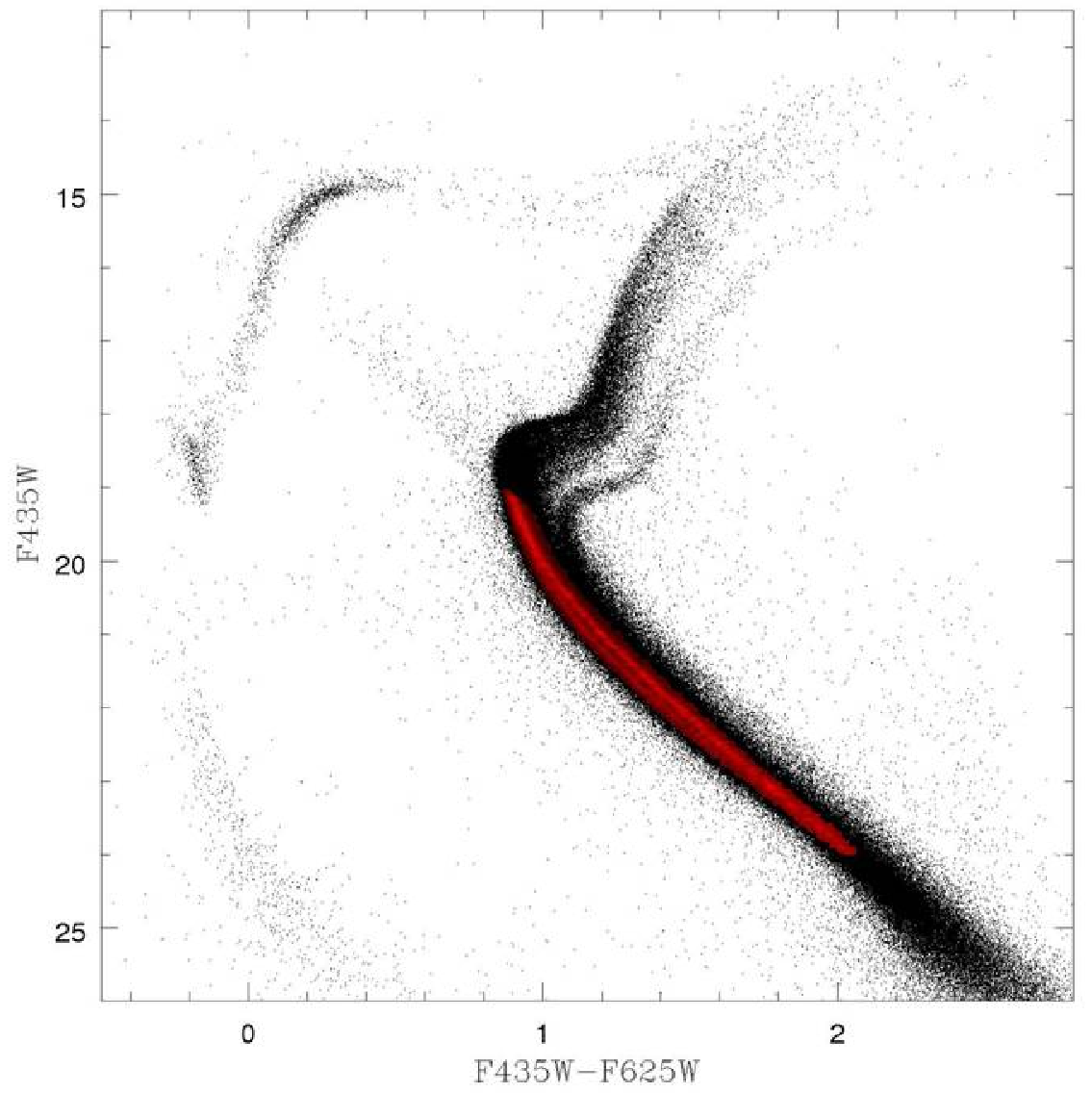} }
\caption[]{The CMD of M54 (left panel) resembles the CMD of $\omega$~Cen in
many aspects. Do the two clusters have had a similar origin?}
\label{fig2}
\end{figure}


The first, direct observational evidence of the presence of more than
one stellar population in a GC was published by Bedin et
al. (2004). Bedin et al. found that, for a few magnitudes below the
turn-off (TO), the main sequence (MS) of $\omega$ Centauri splits in
two (Fig. 1). Indeed, the suspect of a MS split in $\omega$ Cen was
already rised by Jay Anderson in his PhD thesis, but the result was
based on only one external WFPC2 field, and this finding was so
unexpected that he decided to wait for more data and more accurate
photometry to be sure of its reality. Indeed, Bedin et al. (2004)
confirmed the MS split in Jay Anderson field and in an additional ACS
field located 17 arcmin from the cluster center. Now, we know that the
multiple MS is present all over the cluster, though the ratio of blue
to red MS stars diminishes going from the cluster core to its envelope
(Sollima et al. 2007, Bellini et al. 2009, in preparation). We also know that
there is a third, redder MS (see Fig. 1), including about 5\% of the $\omega$~Cen MS stars
(Villanova et al. 2007),
probably related to the most metal rich RGB-a of Pancino et al. (2002).

The more shocking discovery on the multiple populations in $\omega$
Cen, however, came from a follow-up spectroscopic analysis that showed
that the blue MS has twice the metal abundance of the dominant red
branch of the MS (Piotto et al.\ 2005).  The only isochrones that
would fit this combination of color and metallicity were extremely
enriched in helium ($Y\sim 0.38$) relative to the dominant
old-population component, which presumably has primordial helium.

Indeed, the scenario in $\omega$ Cen is even more complex.  As is
already evident in the CMD of Bedin et al. (2004) the three MSs of 
$\omega$~Cen  spread into a highly multiple sub-giant branch
(SGB) with five distinct components characterized by
different metallicities and ages (Sollima et al.\ 2005, Villanova et
al.\ 2007; the latter has a detailed discussion.) 

These results reinforced the suspicion that the multiple
MS of $\omega$ Cen could just be an additional peculiarity of an
already anomalous object, which might not even be a GC, but a remnant
of a dwarf galaxy instead.

On this respect, it might be instructive to compare the CMD of $\omega$~Cen
(Fig 2, right panel), with the CMD of M54 (Fig. 2, left panel). The two CMDs look
rather similar. We know that M54 almost coincides with the nucleus of the 
disrupting Sagittarius dwarf galaxy. And the complexity of the CMD of M54 of
Fig. 2 is indeed due to the fact that we observe, in the same field, both
M54 stars and background/foreground stars of the Sagittarius dwarf nucleus. 
M54 might have originated in the nucleus of its hosting galaxy, or ended there from  elsewhere as a consequence of the dynamical friction (Bellazzini et al. 2008). The important fact here is that M54 and Sagittarius nucleus now are located in the same place, in mutual dynamical interaction. It is very tempting to think that, a few Gyrs ago,
$\omega$~Cen could have been exactly what we now find in the nucleus of the Sagittarius.

The spectacular case of $\omega$~Cen stimulated a number of investigations
which showed that the multiple population scenario is not a peculiarity of a
single object.
Piotto et al. (2007) showed that also the CMD of NGC 2808 is splitted into three
MSs. Because of the negligible dispersion in Fe
peak elements (Carretta et al 2006), Piotto et al. (2007) proposed the
presence of three groups of stars in NGC 2808, with three different He
contents, in order to explain the triple MS. These groups may be associated to the three groups with
different Oxygen content discovered by Carretta et al. (2006). These
results are also consistent with the presence of a multiple, extended HB, as
deeply discussed in D'Antona and Coloi (2004) and D'Antona et
al. (2006). 

Also NGC 1851 must have at least two, distinct stellar populations. In
this case the observational evidence comes from the split of the SGB
in the CMD of this cluster obtained from ACS/HST data (Milone et al. 2008a). 
More recently, (see paper in the present Proceedings), Peter Stetson has 
identified the SGB split also on the cluster envelope, 
thanks to a spectacular photometry  from ground-based wide field images.
Would the magnitude difference between the two SGBs be due only to an age
difference, the two star formation episodes should have been
separated by at least 1 Gyr. However, as shown by Cassisi et
al. (2007), the presence in NGC 1851 of two stellar populations, one
with a normal $\alpha$-enhanced chemical composition, and one
characterized by a strong CNONa anticorrelation pattern could
reproduce the observed CMD split. In this case, the age spread between
the two populations could be much smaller, possibly consistent with
the small age spread implied by the narrow TO of NGC 2808. In other
terms, the SGB split would be mainly a consequence of the metallicity
difference, and only negligibly affected by (a small) age
dispersion. Cassisi et al. (2007) hypothesis is supported by the
presence of a group of CN-strong and a group of CN-weak stars
discovered by Hesser et al. 1982, and by a recent work by Yong and
Grundahl (2007) who find a NaO anticorrelation among NGC 1851 giants.


\begin{figure}[h!]
\begin{center}
{\includegraphics[width=4in]{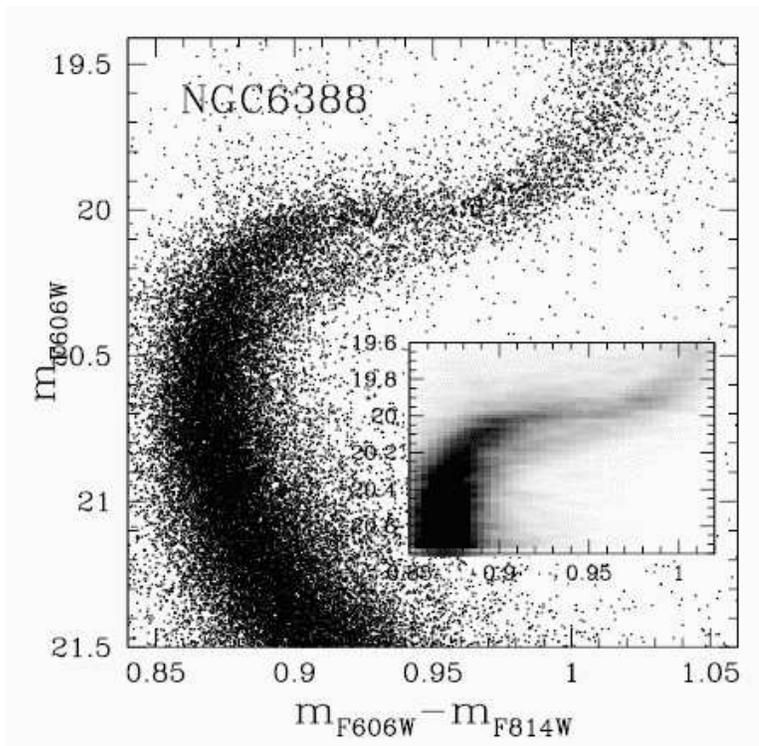}}%
 \caption{The double SGB in NGC 6388.}
   \label{fig3}
\end{center}
\end{figure}

\begin{figure}[h!]
\begin{center}
 \includegraphics[width=4in]{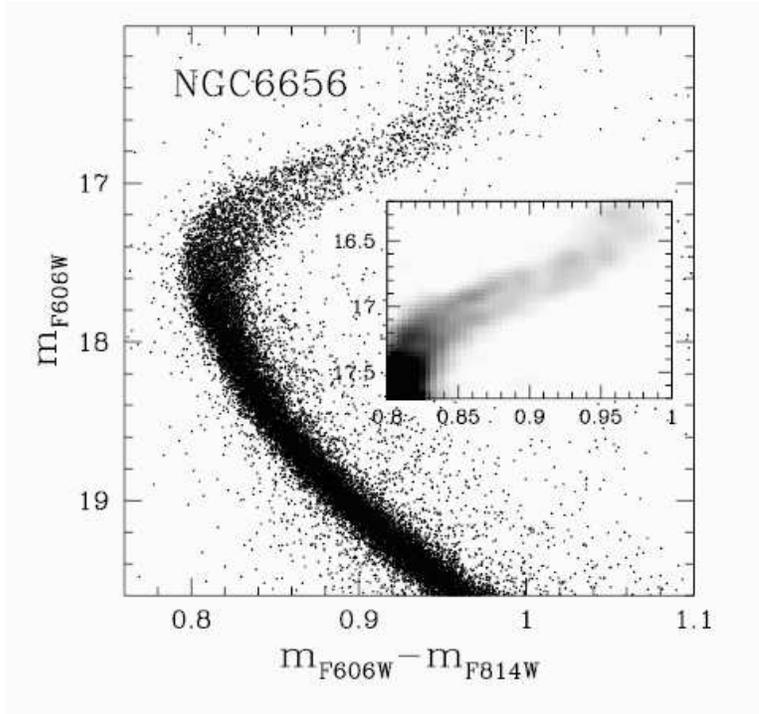} 
 \caption{The double SGB in NGC 6656 (M22).}
   \label{fig4}
\end{center}
\end{figure}

\begin{figure}[h!]
\begin{center}
 \includegraphics[width=4in]{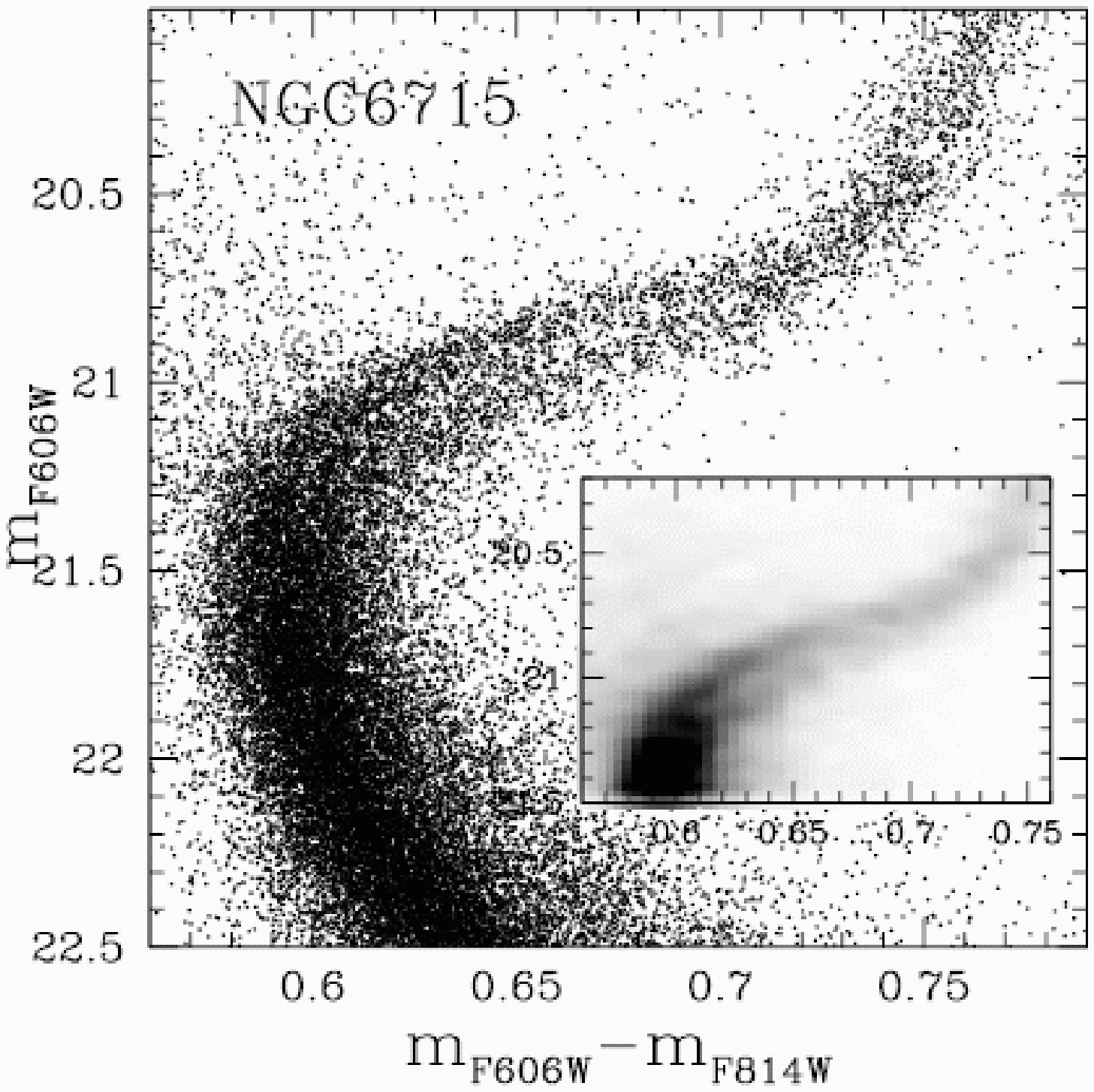} 
 \caption{The double SGB in NGC 6715 (M54).}
   \label{fig5}
\end{center}
\end{figure}

NGC 1851 is considered a sort of prototype of bimodal HB clusters.
Milone et al. (2008a) note that the fraction of fainter/brighter SGB
stars is remarkably similar to the fraction of bluer/redder HB
stars. Therefore, it is tempting to associate the brighter SGB stars
to the CN-normal, s-process element normal stars and to the red HB,
while the fainter SGB should be populated by CN-strong, s-process
element-enhanced stars which should evolve into the blue HB. In this
scenario, the faint SGB stars should be slightly younger (by a few
$10^7$ to a few $10^8$ years) and should come from processed material
which might also be moderately He enriched, a fact that would help
explaining why they evolve into the blue HB. By studying the cluster
MS, Milone et al. (2008a) exclude an He enrichment larger than
$\Delta$Y=0.03), as expected also by the models of Cassisi et
al. (2007). Nevertheless, this small He enrichment, coupled with an
enhanced mass loss, would be sufficent to move stars from the red to
the blue side of the RR Lyrae instability strip.  Direct spectroscopic
measurements of the SGB and HB stars in NGC 1851 are badly needed.

\subsection{More recent findings. I. Galactic GCs}

Prompted by the results on $\omega$~Cen, NGC~2808, and NGC~1851 we 
used HST archive images and new proprietary data (GO10922 and GO11233, PI
Piotto) to search for multiple populations. We are still
working on the optimization of the software for the optimal extraction of
high accuracy photometry from the WFPC2 images (in particular for the
new data acquired in the last months). For the moment, we did not find other cluster
with multiple MSs as in $\omega$~Cen and NGC~2808, thought there are
a couple of suspected cases.

However, we did find many clusters (at least seven, at the moment) with a
double SGB (Piotto et al. 2009, in preparation). Among these, the most 
interesting cases are those of NGC~6388 (Fig. 3), M22 (Fig. 4), and M54 (Fig. 5).

Figure 3 shows that, even after correction for differential reddening, 
the SGB of NGC 6388 closely resembles the SGB of NGC 1851. The results,
originally coming from HST data, have been recently confirmed also using near-IR, multiconiugate adaptive optics images (Moretti et al. 2008) collected with MAD@VLT.
NGC 6388,  as well as its twin cluster NGC 6441, are two extremely peculiar clusters. Since Rich
et al. (1997), we know that, despite their high metal content, higher
than in 47 Tucanae, they have a bimodal HB, which extends to extremely
hot temperatures (Busso et al. 2007), totally un-expected for this
metal rich cluster. NGC 6388 stars also display a NaO anticorrelation
(Carretta et al. 2007). Unfortunately, available data do not allow us
to study the MS of this cluster, searching for a MS split. 
In this context, it is
worth noting that Caloi and D'Antona (2007), in order to reproduce the
HB of NGC 6441, propose the presence of three populations, with three
different He contents, one with an extreme He enhancement of
Y=0.40. Such a strong enhancement should be visible in a MS split, as
in the case of $\omega$ Cen and NGC 2808.  A strong He enhancement and
a consequent MS split may also apply to NGC 6388, because of the many
similitudines with NGC 6441. As soon as the new WF3 and the restored ACS 
instruments at HST will be available, we plan to test these predictions
(GO11739, PI Piotto).

The case of M22 (Fig. 4) is a very interesting one. For decades this cluster
has been suspected to have metallicity variations, including a spread in
[Fe/H]. The iron spread has been controversial till very recently, when,
thanks the UVES@VLT high resolution spectra of RGB stars, 
we (Marino et al. 2009, in preparation)
could demonstrate that, not only the [Fe/H] spread is confirmed, but that,
indeed, there is a bimodal distribution in the iron content, and that this
distribution is correlated with the abundance of s-process elements 
(as Y, Zr, Ba): Stars rich in Fe and Ca are also s-process element rich. The bimodal
distribution of the SGB stars in M22 shown in Fig.~4 might be related to
these two metallicity groups.

Also M54 shows a double SGB (Fig. 5). The fact that the stars
populating the two SGBs are members of the M54 GC, and not field stars,
is confirmed by the fact that they share exactly the same radial 
distribution within our ACS images centered on the cluster center.
Piotto et al. (2009) shows that the double SGB phenomenon is quite
common among massive GCs, though the fraction of stars in the two SGBs varies
from cluster to cluster.

\subsection{More recent findings. II. Magellanic Cloud Clusters}

The multiple population phenomenon in star
clusters is not confined to Galactic GCs only. The suspect that some cluster in the 
Large Magellanic Cloud (LMC)
could host more than one generation of stars has been rised in the past (e.g., Vallenari et al. 1994, Bertelli et al. 2003). However, only when high precision photometry from ACS/HST images became
available, Mackey \& Broby Nielsen 
(2007) could clearly demonstrate the presence of two populations with an age
difference of $\sim$300 Myr in the 2  Gyr old cluster NGC 1846, in the LMC. 
In this case, the presence of the two
populations is inferred by the presence of two TOs in the CMD. Mackey et al. (2008) identified two additional LMC clusters with multiple populations.
More recently, Milone et al. (2008b), from the analysis of the CMDs of 16 intermediate  age LMC clusters using HST archive data, showed that the multiple 
population phenomenon might be rather common among LMC clusters: 11 (70\%!) have CMDs which are
not consistent with the presence of a single, simple stellar population (see
also Kozhurina-Platais et al. poster at this meeting). Also the Small Magellanic
Cloud seems to host a cluster with a CMD not consistent
with a single stellar population (Glatt et al. 2008).

\section{An alternative approach to the search of multiple populations}

The presence of abundance spreads in GC stars is well known. As already
mentioned, in some cases more metallicity groups of stars can be isolated
in the same cluster, as in the case of NGC 2808 (Carretta et al. 2006). 

However, there is a recent finding on the GC M4 
which is worth describing in some details, here. Using high resolution
UVES@VLT spectra of more than 100 giants, Marino et al. (2008) have shown
that also in this cluster, stars show a well defined NaO anticorrelation
(as in all GCs studied so far, searching for this phenomenon). The important
result is that the distribution in Na (or O) content is clearly bimodal
(see inset of Fig. 6),
and this bimodal distribution is correlated with a bimodal distribution
in CN strenght among M4 stars. Stars Na-rich are also CN-strong. 

The bimodality is also visible in the CMDs  built using U images, as shown in Fig. 6. 
This is due to the strong effect of the CN bands on the U magnitude, as demonstrated
by Marino et al. (2008).
This is an interesting feature. Also in the case of M22, where Marino et al. (2009) have
identified two groups of 
stars with different metal contents (see above), the U vs (U-V) CMD allows to distinguish
the two different populations. This RGB split or broadening in the U vs (U-B) CMD may be an
alternative way to search for multiple stellar populations in GCs, in particular for clusters
where large samples of high resolution spectra are not available or not easily obtainable.

\begin{figure}[h!]
\begin{center}
 \includegraphics[width=4in]{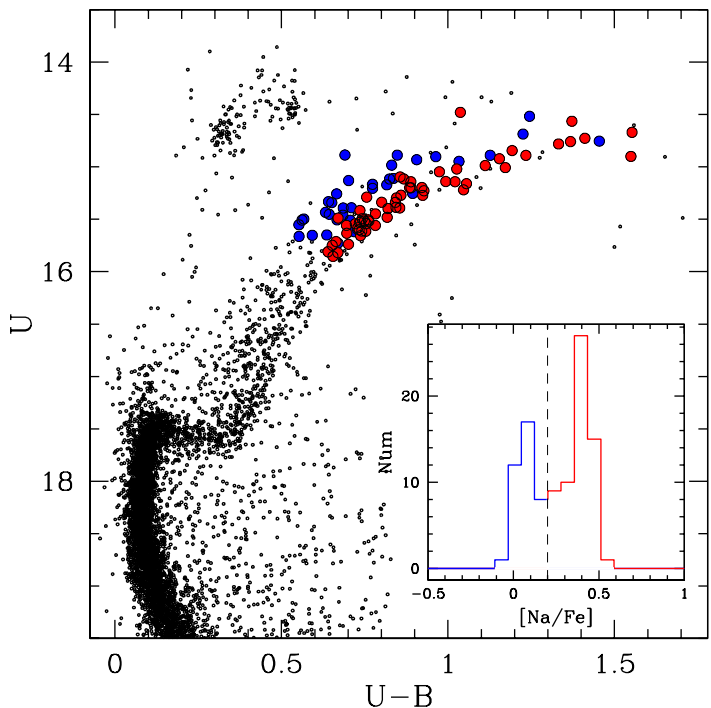} 
 \caption{The Na and O distribution in M4 are bimodal (see inset). The
 bimodal distribution in these chemical elemets reflects also in a bimodal
 distribution of stars along the RGB.}
   \label{fig6}
\end{center}
\end{figure}

\section{Discussion}

In the previous sections we have summarized the direct evidence
we have of multiple generations of stars in star clusters. The observational
scenario is rather complex. So far, multiple populations have been indentified
by the presence of:
\begin{itemize}

\item multiple, distinct  MSs, as in the case of $\omega$~Cen and NGC 2808;

\item TO-SGB splits, in seven Galactic GCs, eleven intermediate age LMC cluster, and
at least one SMC cluster;

\item bimodal or multimodal distribution of light elements, as Na of O, or in the
CN-strength, in some case associated to a broadening or multimodal distribution of
RGB stars in CMD involving the U-band.

\end{itemize}

In general, the multiple population phenomenon differ from cluster to cluster, in the
way it shows itself, in the ratio of the different populations presenting the same cluster, 
in the separation of the different sequences. 
An important property shared by all the clusters where the phenomenon
has been seen so far is that the different populations are distinct. Only in some LMC clusters
there seems to be a broadened distribution, but it is not clear whether
this is just due to photometric errors, which do not allow to separate the single sequences,
or to an intrinsic feature (Milone et al. 2008b). 

At the moment, we cannot say whether the three different manifestations of the
multimodality of cluster stellar populations reflect a single phenomenon.
For example, it has been proposed (Bekki and Mackey 2008) that 
the origin of the bimodal populations in LMC clusters could come from 
an encounter of a young cluster with a giant molecular cloud, where the formation of
a second generation of stars is triggered by the encounter itself. On
the other hand, the multiple populations identified in the (generally more massive) Galactic
globular clusters could be due to a second (or third) generation of stars
which formed from material polluted by the ejecta 
from a variety of possible first generation stars (see Yi review in the present Proceedings).
For sure, these GCs are clearly not simple, single-stellar-population
objects. The emerging evidence is that the star-formation history can
vary strongly from cluster to cluster, and that some GCs are able to produce very
unusual objects, as no such He-rich MS stars have ever been found
elsewhere.  
Reconstruction of this star-formation history requires a a better
understanding of the chemical enrichment mechanisms, but the site of
hot H-burning requested to explain the He enhancement, but also the
NaO anticorrelation remains unclear. There are two requisites: (i)
temperature should be high enough; and (ii) the stars where the
burning occur should be able to give back the processed material to
the intracluster matter at a velocity low enough that it can be kept
within the GC itself (a few tens of km/s).  Candidates include: (i)
Massive ($M>10M_\odot$) rotating stars (Decressin et al.  2007);
(ii) the most massive among the intermediate mass stars undergoing hot
bottom burning during their AGB phase (Ventura et al. 2001), and possibly
(iii) III Population stars (see Yi contribution in the present Proceedings). 
The first two
mechanisms act on different timescales ($10^7$ and $10^8$ yr,
respectively), and both solutions have their pros and cons (Renzini 2008). The
massive star scenario should avoid mixture of O-poor, Na-rich material
with that rich in heavy elements from SNe, while it is not clear how
the chemically processed material could be retained by the
proto-cluster in spite of the fast winds and SN explosions always
associated to massive stars.  Producing the right pattern of
abundances from massive AGB stars seems to require considerable fine
tuning. In addition, both scenarios require that either the IMF of GCs
was very heavily weighted toward massive stars, or that some GCs
should have lost a major fraction of their original population (Bekki
and Norris 2006), and then may even be the remnants of tidally
disrupted dwarf galaxies, as suggested by the complexity in the CMD of
$\omega$ Cen and M54.

The observational scenario is becoming more complex, but, the new
results might have indicated the right track for a comprehensive
understanding of the formation and early evolution of GCs.  We are
perhaps for the first time close to compose what has been for decades
and still is a broken puzzle.

\bigskip

{\bf Acknowledgements.} I wish to warmly thank J. Anderson, Andrea
Bellini, Luigi R. Bedin, Santi Cassisi, Ivan R. King, Antonino P. Milone, Alessia Moretti,
and Sandro Villanova without whom most of the results presented 
in this review would not have been possible. 
A special thanks to Alvio Renzini and Raffaele Gratton for
the many enthusiastic discussions on the subject of multipopulations
in star clusters. I acknowledge partial support by MIUR (PRIN2007) and
ASI under contract ASI-INAF I/016/07/0.

\end{document}